\documentclass[showpacs,preprint,floats,aps,epsfig]{revtex4}
\usepackage{epsfig}

\begin{document}
\title{The Relativistic Rotation}
\author{Z.X. Cao}
\email{caozx@iris.ciae.ac.cn} \affiliation{China Institute of
Atomic Energy, P. O. Box 275(18),
 Beijing, 102413 China
}
\author{Ch.L. Chen}
\author{L. Liu}
\affiliation{ Department of Physics, Beijing Normal University,
Beijing 100875, China }

\date{\today}
%\maketitle

\begin{abstract}
The classical rotation is not self-consistent in the framework of
the special theory of relativity. the Relativistic rotation is
obtained, which takes the relativistic effect into account. It is
demonstrated that the angular frequency of classical rotation is
only valid in local approximation. The properties of the
relativistic rotation and the relativistic transverse Doppler
shift are discussed in this work.

4.3, 2003 at 12:58

%\textbf{keyword} the relativistic rotation, the special theory of
%relativity, classical rotation, angular frequency, transverse
%Doppler effect

\end{abstract}

\maketitle
%\maketitle

\section{Introduction}

The special and general theory of relativity which was developed
in the beginning of the twentieth century, especially through
Einstein's work\cite{ein1,ein2,ein3}, has its roots far back in
the past. But the rotational reference systems utilized in the
present researches of physics are all classical\cite{mo1,ros1},
although there exists the special and general theory of
relativity. In the classical rotation, the velocity $v$ of one
point relative to the center of circle is proportional to its
distance $r$ from the center of circle and its angular frequency
$\omega_0$. However, this relationship is not suitable within the
special relativistic theory.

The velocity formula of classical rotation is given as
$v=r\omega_0 $. It could be understood from the following three
aspects. First, the foundation of experiments, i.e., the
experience of daily life. Second, Galilean transformations
\cite{mo1}. Third, the general conclusion derived from the first
and the second points. Following the inference of Galilean
transformations, the velocity formula of classical rotation is
generally come into existence under any rotational conditions,
such as the condition of long-distance and high-velocity. It means
that the formula $v=r\omega_0 $ is self-consistent in classical.

However, the application of Galilean transformations in
high-velocity condition is proved early to be not
reliable\cite{mo1,ros1}. It is not correct for the object with
high velocity  and must be replaced with the principle that could
embodies the relativistic properties. Therefore, the extending
velocity formula $v=r\omega_0 $ of classical rotation to arbitrary
is suspicious, and it is need to reevaluate. We should start from
the special relativistic theory to reconsider the rotation with
high velocity.

the proper relativistic metric of rotational reference system
should be confirmed. The four-dimension metric of the uniform
rotational system is classical in the general theory of relativity
now\cite{mo1,ros1}. It means the proper relativistic present of
the rotational system couldn't be given in the general theory of
relativity. And the uniform rigid rotation is instantaneous and
local inertial, so we could research it in the frame of the
special theory of relativity.

To perform the relativistic analysis of uniform rotation, the
Galilean transformations and its deduced conclusion, i.e., the
formula $v=r\omega_0 $ is extended to be correct in infinite
distances, should be discarded. Nevertheless, we hold the opinion
as usual experience that the classical formula of rotation is
applicable within short distances. In every observational point,
what the observer measures is just the rotational velocities in
very close distances, while for the rotational velocities far away
from the observers, the calculation using the transformation of
velocity \cite{mo1} in the special theory of relativity is needed.

We analysis the classical rotation firstly, and point out its
non-self-consistent disadvantage. A rotational formula of angular
velocity with relativistic character is obtained, which approach 
to classical velocity in the limit of non-relativity. It shows
that the classical rotation is only local and slow-velocity
approximation of the relativistic one.

\section{The Relativistic Rotation}
In classical rotation the velocity $V$ of one point relative to
another point is equal to the product of its angular frequency
$\omega _{0}$ and its distance from center point $R$.

\begin{equation}\label{eq1}
V =R\omega _{0}
\end{equation}

\begin{figure}[ht]
\centerline{\hspace{-0.in}
\epsfig{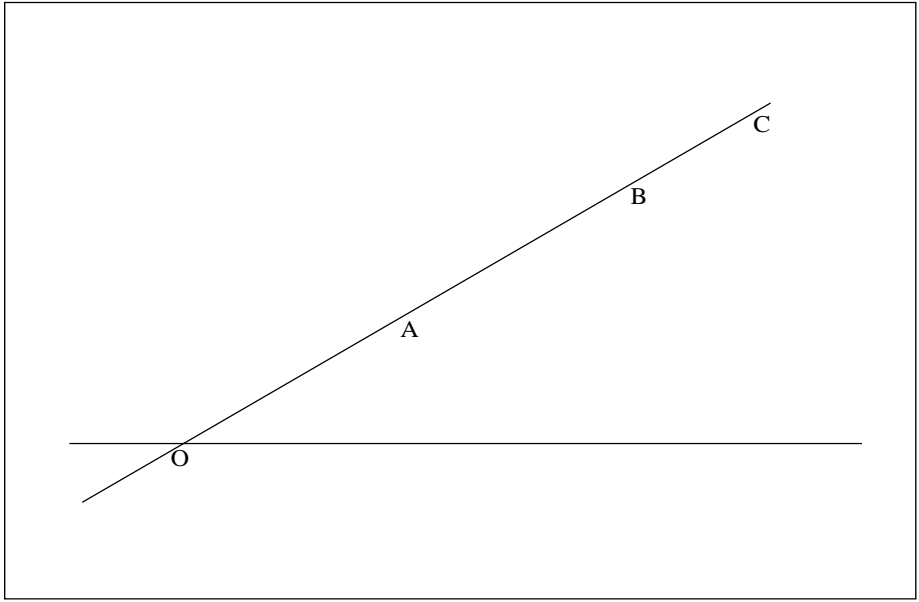}}
%\vspace{0.2in}
\caption{The sketch map of the rotating pole} \label{fig1}
\end{figure}

We show in Fig.\ref{fig1} the rotation of a long pole around point
$O$ with frequency $\omega _{0}$. We assume that the pole is in a
part of space so far from all masses that all gravitational
effects can be neglected. The distance of point $A$ and $B$ from
center point $O$ is $R$ and $2R$, respectively. Their velocities
relative to the center point $O$ are $V_{AO}=R\omega _0$ and
$V_{BO}=2R\omega _0$, respectively, in the classical limit.
However, if one considers the transformation formula of the
velocity in the special theory of relativity, the velocity of
point $B$ over point $O$ becomes
\begin{equation}\label{eq2}
V_{BO}^{\prime }=\frac{2R\omega _{0}}{1+(R\omega _{0}/c)^{2}}\neq
2R\omega _{0}.
\end{equation}

We can see from formula (\ref{eq2}) that the velocity obtained
from classical limit can not coincide with the one derived from
the relativistic one due to the use of the transformation formula
of the velocity in the special theory of relativity. It implies
that the classical rotation is not self-consistent in the
framework of the special theory of relativity.

No one will suspect the accuracy of classical rotation formula in
daily life. In fact, the correction term in the denominator of
Eq.2 is nearly equal to zero under the ordinary condition.
Nevertheless, when the classical rotation formula is extended to
infinite, i.e. $R\omega _{0}\approx c$, the Galilean
transformations is still applied but not the special relativity.
It is shown that this kind of formula can not fulfill the special
theory of relativity. This motivates us to improve the classical
rotation.

because the general theory of relativity only gives the principle
of equivalence, has nothing to do with the rotational
system\cite{mo1,ros1}, the relativistic rotational system can not
be obtained through it. Therefore, we start from the special
theory of relativity to study the rotational system, and to derive
the kinematical equation of the relevant rotational system. From
the kinematical point of view, the investigation of the rotation
based on the special theory of relativity is suitable.

We can see from formula (\ref{eq2}) that the velocity
(Eq.\ref{eq1}) of classical rotation is not correct for the whole
description of rotation under the framework of special relativity.
In order to obtain the suitable expression of the rotational
velocity, we discard the assumption of formula (\ref{eq1}) for the
whole-description and only consider it as the local approximation.
The formula (1) is only suitable in the limit of $R\omega _0 \ll
c$. Furthermore, we think that there is not intrinsic discrepancy
between the inertial and noninertial system if the gravitation is
not taken into account, and the instantaneous localization of
rotational system is approximately regarded as the inertial one.
Moreover, the two adjacent systems of instantaneous localization
fulfill the principle of special relativity.

Taking point $A$ as the reference point, we would determine the velocity of
point $B$ relative to point $A$ in the framework of special relativity using
the former conditions.

In Fig.\ref{fig1} the observers at each point would see that the
object in very small area rotates uniformly around the point.
Assuming that the distance between point $A$ and $B$ is $R$, we
now calculate the velocity of point $B$ relative to point $A$ at
the rotating long-pole. Obviously, we could not measure the
velocity at point B directly. However, it could be measured via
the following method: the distance between point $A$ and $B$ is
divided into $n$ equal parts, and each point in every equal parts
could be considered as one instantaneous inertial reference
system. Therefore, the velocity of the point, where the distance
from the original point is $\Delta R=R/n$, could be measured, and
the velocity of this point is $V_1=\Delta R \omega _0$. Then, we
would measure the next $\Delta R$ position, where the distance
from the original point is $2\Delta R$, and the velocity of
position $ 2\Delta R$ relative to the position $\Delta R$ is also
$\Delta R \omega _0$. The measurement is continued until reaching
the position of $R$, and then, we could calculate the velocity of
the last point relative to the beginning point taking advantage of
the relativistic velocity transfer formula. Based on the relations
of velocity in these instantaneous inertial system, the velocity
$V_n$ is the very relativistic velocity $V_{re}$ when $n$
approaches infinite.

The velocity of each point relative to point $A$ is
\begin{equation}\label{eq3}
\begin{array}{rcl}
\Delta R\ \ \ \ \ \ V_{1} & = & \Delta R\omega _{0} \\
  2\Delta R\ \ \ \ \ \ V_{2} & = & \frac{\Delta R\omega
_{0}+V_{1}}{1+V_{1}\Delta R\omega _{0}/c^{2}} \\
  3\Delta R\ \ \ \ \ \ V_{3} & = & \frac{\Delta R\omega
_{0}+V_{2}}{1+V_{2}\Delta R\omega _{0}/c^{2}} \\
  \cdots &   &   \\
n\Delta R=R\ \ \ \ \ \ V_{n} & = & \frac{\Delta R\omega
_{0}+V_{n-1}}{ 1+V_{n-1}\Delta R\omega _{0}/c^{2}}
\end{array}
\end{equation}

The velocity of inertial system $B$ relative to the one of $A$ is
the value of $V_{n}$ in the limit of $n$ approaching infinite.
According to formula (\ref{eq3}), we obtain
\begin{equation}\label{eq4}
\begin{array}{rcl}
V_{n} &=& \frac{c^{2}}{\Delta R\omega _{0}}\frac{V_{n-1}\Delta
R\omega _{0}+1+(\Delta R\omega _{0})^{2}/c^{2}-1}{1+V_{n-1}\Delta
R\omega _{0}/c^{2}}\\
 &=& \frac{c^{2}}{\Delta R\omega _{0}}\left[
1-\frac{1-(\Delta R\omega _{0})^{2}/c^{2}}{1+V_{n-1}\Delta R\omega
_{0}/c^{2}}\right]
\end{array}
\end{equation}
Defining $\eta =1-(\Delta R\omega _{0})^{2}/c^{2}$,
$U_{i}=\frac{V_{i}\Delta R\omega _{0}}{c^{2}}$ we could rewrite
formula (4) as
\begin{equation}\label{eq5}
U_{n}=1+\frac{-\eta }{1+U_{n-1}},\cdots ,U_{2}=1+\frac{-\eta
}{1+U_{1}},U_{1}=1-\eta
\end{equation}

The $U_{n}$ could be expressed in terms of continued fraction
\begin{equation}\label{eq6}
U_{n}=1+\frac{-\eta }{2+}\frac{-\eta }{2+}\frac{-\eta }{2+} \cdots
\frac{-\eta }{2-\eta }
\end{equation}

The recursive properties of the continued fraction is as
follow\cite{har1,roc1}
\begin{equation}\label{eq7}
\begin{array}{ccc}
 &U_{n,1}=\frac{p_{1}}{q_{1}}=\frac{1}{1},U_{n,2}=\frac{p_{2}}{
q_{2}}=\frac{2-\eta}{2},\cdots ,& \\
&U_{n,i}=\frac{p_{i}}{q_{i}},\cdots
,U_{n}=U_{n,n}=\frac{p_{n} }{q_{n}}& \\
  &p_{i}=2p_{i-1}-\eta p_{i-2},q_{i}=2q_{i-1}-\eta q_{i-2}\ \
(i=3,4,\cdots ,n)&
\end{array}
\end{equation}
where the series of $p_{i}$ and $q_{i}$ could be solved in terms
of the characteristic equation of recursive series\cite{ric1}

\begin{equation}\label{eq8}
x^{2}-2x+\eta =0
\end{equation}
The two solutions of formula (\ref{eq8}) are
\begin{equation}\label{eq9}
x_{1}=1+\sqrt{1-\eta },\ x_{2}=1-\sqrt{1-\eta }
\end{equation}
Then, the general solutions of $p_{i}$ and $q_{i}$ are
\begin{equation}\label{eq10}
\begin{array}{ccc}
p_{i} &=& A_{1}(1+\sqrt{1-\eta })^{i}+A_{2}(1-\sqrt{1-\eta })^{i} \\
 q_{i} &=& A_{3}(1+\sqrt{1-\eta })^{i}+A_{4}(1-\sqrt{1-\eta
})^{i}
\end{array}
\end{equation}
It is obtained with the original condition of formula (\ref{eq7}).
\begin{equation}\label{eq11}
\begin{array}{ccc}
&A_{1}=\frac{\sqrt{1-\eta }}{2},A_{2}=-\frac{\sqrt{1-\eta }}{2}&
\\
&A_{3}=\frac{1}{2},A_{4}=\frac{1}{2}&
\end{array}
\end{equation}
Inserting formula (\ref{eq11}) into formula (\ref{eq10}) and
considering the formula (\ref{eq7}), we can get
\begin{equation}\label{eq12}
U_{n}=\sqrt{1-\eta }\frac{(1+\sqrt{1-\eta })^{n}-(1-\sqrt{1-\eta })^{n}}{(1+%
\sqrt{1-\eta })^{n}+(1-\sqrt{1-\eta })^{n}}
\end{equation}

Inserting $\eta =1-(\Delta R\omega _{0}/c)^{2}$ and $\Delta R=R/n$
into the former formula (\ref{eq12}), the relativistic velocity in
the limit of $n\rightarrow \infty $ is
$$\frac{V_{re}}{c}={\lim \limits _{n\rightarrow \infty }}\frac{U_{n}}{\sqrt{%
1-\eta }}={\lim \limits_{n\rightarrow \infty }}\frac{(1+\frac{
R\omega _{0}}{nc})^{n}-(1-\frac{R\omega _{0}}{nc})^{n}}{(1+\frac{%
R\omega _{0}}{nc})^{n}+(1-\frac{R\omega _{0}}{nc})^{n}}=\tanh (%
\frac{R\omega _{0}}{c})$$ i.e.,
\begin{equation}\label{eq13}
V_{re}(R)=c\tanh (\frac{R\omega _{0}}{c})
\end{equation}

which is the expecting formula of relativistic rotation velocity.
We calculate numerically the results of formula (\ref{eq3}), and
the consistent results are reached with formula (\ref{eq13}).
Comparing with the classical rotation formula (\ref{eq1}), We can
see that the classical rotation is the approximation of formula
(\ref{eq13}) in the limit of $ c\rightarrow\infty$. For the case
far from classical rotation, an additional correction term
$-R^{3}\omega _{0}^{3}/c^{2}$ is obtained from formula (13) in
comparison with formula (1), i.e.,
\begin{equation}\label{eq14}
V_{re}\simeq R\omega _{0}-R^{3}\omega _{0}^{3}/c^{2}
\end{equation}

Considering the three points ($A$, $B$, and $C$) in the same line,
we could calculate easily the their velocity using the
relativistic transfer formula of velocity and formula (\ref{eq13})
\begin{equation}\label{eq15}
V_{CA}=\frac{V_{CB}+V_{BA}}{1+V_{CB}V_{BA}/c^{2}}=c\tanh (\frac
{R_{CA}\omega _{0}}{c})
\end{equation}

These show that the velocity relation in formula (\ref{eq13}) is
self-consistent in the framework of special relativity, and now
the relativistic space-time interval are different with the
classical one\cite{mo1}. It could be expressed in cylindrical
coordinates
\begin{equation}\label{eq16}
ds^{2}=dr^{2}+dz^{2}+r^{2}d\phi ^{2}+2cr\tanh (r\omega
_{0}/c)d\phi dt-c^{2}\cosh^{-2}(r\omega _{0}/c)dt^{2}
\end{equation}

It is also observed from the comparison between formula
(\ref{eq1}) and (\ref{eq13}) that the angular velocity in the
classical rotation is only the value in token of the frequency of
relativistic rotation, while not the real angular velocity of
relativistic rotation. The relativistic angular velocity $\omega$
of two points in relativistic rotation is no longer the physical
invariable in the classical rotation. It is with respect to not
only the token angular velocity but also the distance of the two
points, i.e.,
\begin{equation}\label{eq17}
\omega (r)=\frac{c\tanh (r\omega _{0}/c)}{r}
\end{equation}

It is obviously $\omega (r)\rightarrow \omega _{0}$ with
$r\rightarrow 0$ and $\omega (r)\rightarrow 0$ with $r\rightarrow
\infty $, which is different with the classical one. In classical
rotation reference systems, what could be observed is only within
the range from observation point extending to $c/\omega _{0}$.
Otherwise, with larger distances, the observation velocity will
greater than velocity of light that could not be achieved by
realistic objects. Nevertheless, in the relativistic rotation
reference systems, the former case could not exist. It could be
found from formula (\ref{eq13}) that all the velocity are less
than velocity of light $c$. Furthermore, the discrepancy of
angular velocity with the relativistic uniform rotation are
generally not observed in the dailylife due to the condition
$R\omega _{0}\ll c$.

\section{Transverse Doppler effect in relativistic rotation}

The expression of transverse Doppler effect can be obtained from
special relativity in classical rotation \cite{ros1}. For
instance, there is a light-source with frequency $\nu_0$ at point
B in Fig.\ref{fig1}. Following the Doppler formula in special
relativity and formula (\ref{eq1}), the frequency of the light at
point A in classical rotation is given as
\begin{equation}\label{eq18}
\nu _{classic}=\nu _{0}\sqrt{1-(R\omega _{0}/c)^{2}}
\end{equation}

However, in relativistic rotation, through the Doppler formula in
special relativity and formula (\ref{eq13}), the observed
frequency of the light at point A is as follow
\begin{equation}\label{eq19}
\nu _{relativity}=\nu _{0}\sqrt{1-\tanh ^{2}(R\omega _{0}/c)}=\nu
_{0}\cosh^{-1}(R\omega _{0}/c)
\end{equation}
\begin{figure}[ht]
\centerline{\hspace{-0.5in}
\epsfig{file=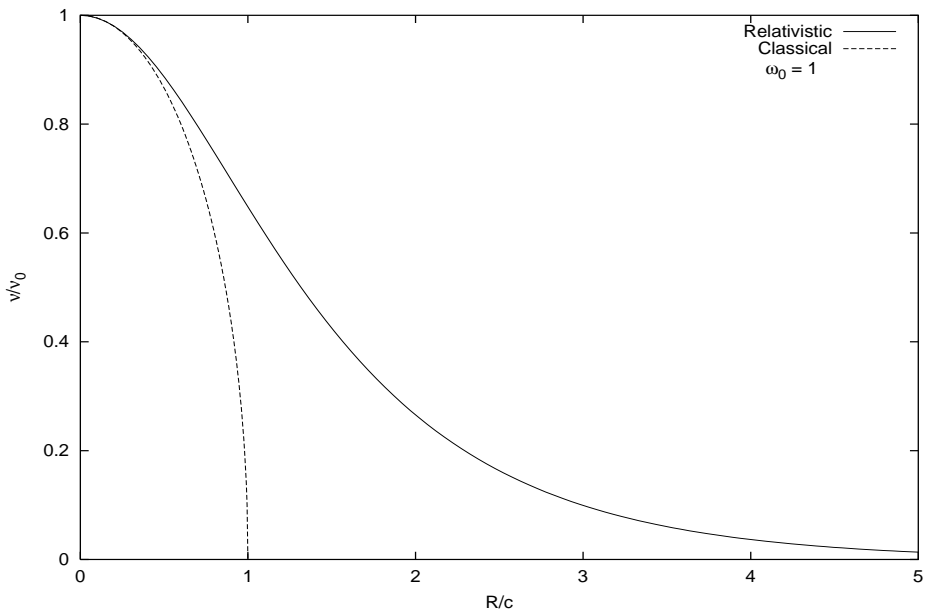,width=3.15in,height=4.8in,angle=270}}
%\vspace{0.2in}
\caption{The transverse Doppler shift in the classic and
relativistic rotation} \label{fig2}
\end{figure}
For further understanding, we plot in Fig.\ref{fig2} the
transverse Doppler shift for classical and relativistic rotations,
respectively, for the comparison between formula (\ref{eq18}) and
(\ref{eq19}). The product $R\omega _0$ in classical rotation can
not greater than velocity of light, otherwise, imaginary number
will occur in the frequency. However, in relativistic rotation,
this product may greater than the velocity of light, and the
properties of $\nu_{relativity} $ in equation (\ref{eq19}) will
not be affected. The difference of these two cases would only
exhibit when the value of $R\omega _0$ approaches the velocity of
light. Expanding the series of formula (\ref{eq18}) and
(\ref{eq19}) to their order with $R^4$, We can get their
difference
\begin{equation}\label{eq20}
\Delta \nu =\nu _{relativity}-\nu _{classic}=\nu_0\frac{(R\omega _{0}/c)^{4}}{3}%
+O(R^{4})
\end{equation}
It is obviously showed in formula (\ref{eq20}) that the
discrepancy is hardly observed in ordinary conditions. However, we
may observe this discrepancy in the large-scale rotations such as
the rotation of galaxy.

\section{summary}

The classical rotation which considers the rotation formula
(\ref{eq1}) as the whole-conditions is not self-consistent in the
special relativistic description. We propose the velocity
expression (\ref{eq13}) of relativistic rotation taking advantage
of formula (\ref{eq1}) only as the local condition. This velocity
is self-consistent for the whole description of rotation. The
relativistic rotation can reduce to the classical one in the
classical conditions, while greatly different properties are
exhibited in comparison with the classical rotation in the
relativistic condition, i.e. $R\omega _0 \sim c$. With the
flourish development of astrophysics, the self-consistent
relativistic rotation would make one understand the phenomena of
celestial rotations more suitably than the classical one.
Moreover, the relativistic rotation may provide a feasible way for
the basic and theoretical description of rotation.

\end{document}